\documentstyle[12pt]{article}

\def\MeV{\,{\rm MeV}}

\def\cmm2{{\,\rm cm^{-2}}}
\def\cm2{{\,{\rm cm}^2}}
\def\cmm3{{\,{\rm cm}^{-3}}}
\def\gcmm3{{\,{\rm g\,cm^{-3}}}}

\def\la{\mathrel{\mathpalette\fun <}}

\def\fun#1#2{\lower3.6pt\vbox{\baselineskip0pt\lineskip.9pt
  \ialign{$\mathsurround=0pt#1\hfil##\hfil$\crcr#2\crcr\sim\crcr}}}

\begin{document}
\baselineskip=16pt
\pagestyle{empty}
\begin{center}
\rightline{FERMILAB--Pub--92/298-A}
\rightline{October 1992}
\rightline{submitted to {\it Physical Review D}}
\vspace{.2in}

{\bf EFFECT OF NEUTRINO HEATING ON PRIMORDIAL NUCLEOSYNTHESIS}\\
\vspace{.2in}

Brian D. Fields,$^{1}$ Scott Dodelson,$^2$ and Michael S. Turner$^{1,2,3}$ \\
$^1${\it Department of Physics,\\
The University of Chicago, Chicago, IL  60637-1433}\\
$^2${\it NASA/Fermilab Astrophysics Center,
Fermi National Accelerator Laboratory, Batavia, IL  60510-0500}\\
$^3${\it  Department of Astronomy \& Astrophysics,\\
Enrico Fermi Institute,
The University of Chicago, Chicago, IL 60637-1433}\\
\end{center}

\vspace{.3in}

\centerline{\bf ABSTRACT}
\bigskip

\noindent We have modified the standard code for
primordial nucleosynthesis to include the
effect of the slight heating of neutrinos by $e^\pm$ annihilations.
There is a small, systematic change in the $^4$He yield,
$\Delta Y \simeq +1.5\times 10^{-4}$, which is insensitive
to the value of the baryon-to-photon ratio $\eta$
for $10^{-10}\la \eta \la 10^{-9}$.
We also find that the baryon-to-photon ratio decreases
by about 0.5\% less than the canonical factor of 4/11
because some of the
entropy in $e^\pm$ pairs is transferred to neutrinos.
These results are in accord with recent analytical estimates.

\newpage

\pagestyle{plain}

\setcounter{page}{1}

\newpage

\section{Introduction}


The concordance between the predictions of primordial
nucleosynthesis and the observed abundances of D, $^3$He,
$^4$He, and $^7$Li is one of the cornerstones
of the hot big-bang cosmology,
and provides its earliest test.  Because of this and
the great interest in the very early history of the Universe,
primordial nucleosynthesis
has been called ``the gateway to the early Universe.''
Further, big-bang nucleosynthesis has been exploited to provide
the most accurate determination of the baryon density \cite{ytsso}
and to probe particle physics, e.g., the stringent limit
to the number of light neutrino species \cite{nulimit}.


Over the past decade there has been continued
scrutiny of primordial nucleosynthesis, both on
the theoretical side and on the observational side:
Reaction rates have been updated and the effect
of their uncertainties quantified \cite{rates},
finite-temperature corrections have been taken into
account \cite{dicusetal}, and the effect of inhomogeneities in the
baryon density explored \cite{matthews}; the primordial abundance
of $^7$Li has been put on a firm basis \cite{lithium}, the production
and destruction of D and $^3$He have been studied carefully
\cite{dhe3}, and astrophysicists now argue about the third significant
figure in the primordial $^4$He abundance \cite{he4}.  The
result is that the ``concordance region''
of parameter space has continued
to shrink.  The predicted and measured primordial
abundances agree provided:  the baryon-to-photon ratio
lies in the narrow interval $3\times 10^{-10}\la \eta \la
5\times 10^{-10}$ and the equivalent number of light
neutrino species $N_\nu \la 3.4$ \cite{walker}.
The trend motivates the study of smaller and smaller effects,
and, in particular, the present examination
of the small effect of the heating of neutrinos
by $e^\pm$ annihilations.


To place our work in perspective, let us enumerate the
usual assumptions underlying primordial nucleosynthesis:
(i)  Friedmann-Robertson-Walker cosmology; (ii)
the input of various nuclear
reaction cross sections, the most important of which
is the matrix element for the processes that interconvert
neutrons and protons; (iii) $N_\nu$
nondegenerate neutrino species, i.e., neutrino chemical
potentials $|\mu_\nu| \ll T$;
and (iv) the complete decoupling
of neutrinos from the electromagnetic plasma
before the entropy in $e^\pm$ pairs is
transferred to photons.  It is the final assumption
that our work addresses.

It has long been known that neutrino interactions with
the electromagnetic plasma, e.g., $\nu +{\bar\nu} \leftrightarrow
e^+ + e^-$, $\nu + e^\pm \leftrightarrow \nu + e^\pm$, and so on,
become ineffective (interaction rate per particle $\Gamma$
less than the expansion rate $H$) at a temperature of the order
of a few $\MeV$.  Since $e^\pm$ pairs do not disappear
and transfer their entropy to the plasma
until a temperature of the order of
$m_e/3\sim 0.1\MeV$, one expects that neutrinos
do not share in the $e^\pm$ entropy transfer.  It then follows
that long after the $e^\pm$ pairs disappear the ratio of the photon
and neutrino temperatures should be $(11/4)^{1/3}$
and the baryon-to-photon ratio should decrease
by 4/11 \cite{Weinberg}.


Neither neutrino decoupling nor
the disappearance of $e^\pm$ pairs are instantaneous events, and so
one might expect neutrinos to share slightly in the $e^\pm$ entropy transfer
and to have a higher temperature than $(4/11)^{1/3}T_\gamma$.
Because the yields of primordial nucleosynthesis are very
sensitive to the neutron fraction, which
around the time of nucleosynthesis is determined by the
rate of neutron-proton interconversions through the processes
$n+e^+ \leftrightarrow p+{\bar \nu}_e$, $n+\nu_e \leftrightarrow
p+e^-$, and $n\leftrightarrow p+e^- +{\bar\nu}_e$,
they depend critically upon the neutrino temperature
as the rates for these processes vary as
$T_\nu^5$.  Even a slight amount of neutrino heating is potentially
important for the $^4$He since its abundance
is now discussed to three significant figures.

A number of authors have tried to quantify
neutrino heating and its effect on the $^4$He yield
\cite{dicusetal,hh,rm,df,dt}.  With the except of the
most recent work, Refs. \cite{df,dt},
previous estimates were ``one-zone'' calculations,
i.e., the integrated perturbation to the neutrino
energy density $\delta\rho_\nu$ was calculated
rather than the perturbations to the neutrino
phase-space distribution functions.  All authors agree
that neutrino heating of the electron neutrinos increases
their energy density by about 1\% (slightly less for
$\nu_\mu$ and $\nu_\tau$, as they only have neutral-current
interactions), and with one exception, Ref. \cite{rm},
all estimate the change in
the mass fraction of $^4$He synthesized to be of the
order of $\Delta Y \sim 10^{-4}$, though there is no consensus
as to the sign of this small change; the authors of Ref.
\cite{rm} estimate the change to be 30 times larger, $\Delta Y
\simeq -0.003$.


In this work we incorporate the results of the
most detailed treatment of neutrino heating \cite{dt}
into the standard big-bang nucleosynthesis code \cite{kawano}.
For the interesting range of the baryon-to-photon
ratio we find a systematic increase in the $^4$He abundance
of $\Delta Y = 1.5\times 10^{-4}$, very close
to the semi-analytical estimate in Ref. \cite{dt}; we find
similar fractional changes for the abundances
of the other light elements.  By integrating the first law of thermodynamics
we find that due to neutrino heating
the baryon-to-photon ratio decreases by about 0.5\% less
than the canonical factor of 4/11; again,
in good agreement with the estimate of Ref. \cite{dt}.

We trace the discrepancy in the predicted sign of the change
in $^4$He yield to other authors not considering
all of the effects of neutrino heating on
the $^4$He yield.  In order to check the interesting
claim that $\Delta Y = -0.003$
we have also modified the nucleosynthesis
code to take into account the effect of neutrino heating
as computed in Ref. \cite{rm}; however, we find that
the predicted change in $^4$He is only $\Delta Y = +1.1\times
10^{-4}$, which is consistent with the more detailed
treatment of neutrino heating.  Since the authors of Ref. \cite{rm}
give few details concerning
the changes they made in the nucleosynthesis code,
it is not possible to explain this discrepancy, though
we are very confident that the change is not as large as they state.

Our paper is organized as follows; in the next Section
we discuss the changes that must be made in the nucleosynthesis
code when neutrino heating is taken into account
and how we implemented them.  In the final Section, we
discuss our numerical results, compare them
to previous estimates for the change in $^4$He production,
and finish with some concluding remarks.

\section{Modifications to the Standard Code}

\subsection{Role of neutrinos}

The slight heating of neutrinos by $e^\pm$ annihilations
causes:  (i) small perturbations to the neutrino
phase-space distributions; and (ii) small decrease in the temperature
of the electromagnetic plasma (at fixed value of
the cosmic scale factor $R(t)$) since neutrinos
take energy away from the electromagnetic plasma.
To understand how these changes affect
the outcome of nucleosynthesis,
let us first review how neutrinos ``participate''
in primordial nucleosynthesis.

Neutrinos play several roles; first,
in governing the neutron-to-proton ratio.
Specifically, the electron neutrino and
antineutrino phase-space distributions affect
the rates (per nucleon) for the reactions
that interconvert neutrons to protons and vice versa,
$\lambda_{np}$ and
$\lambda_{pn}$.  In the standard treatment these rates are computed
by integrating the well known tree-level matrix element
squared over the appropriate (thermal) Fermi-Dirac distributions
(see e.g., Refs. \cite{dicusetal,Weinberg}).
Because of neutrino heating, the electron-neutrino distribution is
given by the usual thermal part plus a small perturbation,
which results in small changes to the weak rates,
$\delta \lambda_{pn}$ and $\delta \lambda_{np}$.

The other roles neutrinos play involve their
contribution to the energy density of the Universe.
The total energy density determines the expansion
rate of the Universe:
\begin{equation}
H^2\equiv \left({{\dot{R}}\over{R}}\right)^2 =
{8 \pi G \rho_{\rm TOT}\over 3},
\end{equation}
where $\rho_{\rm TOT} = \rho_{\gamma} + \rho_e +
\rho_\nu + \rho_B$.  Because of rapid electromagnetic
interactions the electromagnetic plasma is always
in thermal equilibrium so that $\rho_{\rm EM}(T_\gamma )
\equiv \rho_\gamma +\rho_e$ is only a function of
the photon temperature $T_\gamma$.  And of course,
the baryonic contribution to the energy density, $\rho_B$,
is very tiny as the Universe at this early time is
radiation dominated.

In the absence of neutrino heating by $e^\pm$ annihilations
the neutrino temperature just red shifts
with the expansion, $T_{0\nu} \propto R^{-1}$, and
the neutrino energy density $\rho_{0\nu} \propto {1 /{R^4}}$.
When neutrino heating is taken into account
\begin{equation} \label{eq:zero}
\rho_{\nu} = \rho_{0\nu}+\delta \rho_{\nu},
\end{equation}
where $\rho_{0\nu}$ is the energy density in all
three neutrino species in the absence of neutrino
heating, and $\delta\rho_\nu$ is the sum over all
three species of the additional energy density
due to neutrino heating.  In Ref. \cite{dt} the evolution of the
perturbation to the phase-space distribution of each species is
computed; for electron neutrinos $\delta\rho
/\rho$ approaches about 1.2\%, and for $\mu$
or $\tau$ neutrinos about 0.6\%; thus,
$\delta\rho_\nu /\rho_\nu$, the average over the three species,
is about 0.7\%.

The neutrino energy density also appears
in the first law of thermodynamics, which
governs the rate at which the photon temperature
decreases with time:
\begin{equation} \label{eq:first}
d[\rho_{\rm TOT}V] = -p_{\rm TOT}dV,
\end{equation}
where the total pressure $p_{\rm TOT}=p_{\rm EM}
+p_\nu +p_B$, $p_{\rm EM}(T_\gamma )=p_\gamma + p_e$,
 and $V=R^3$.  Since we shall
assume that the three neutrino species are
very light, $m_\nu \ll 1\MeV$, the neutrino
pressure $p_\nu = \rho_\nu /3$.  In the absence
of neutrino heating the neutrino energy density
drops out of Eq. (\ref{eq:first})
equation since $\rho_{0\nu} \propto R^{-4}$.
When neutrino heating is taken into account this
is no longer true; as we shall see, the additional
term in this equation involving $\delta\rho_\nu$
leads to a ``back reaction'' resulting in a slight
cooling of the electromagnetic plasma.

\subsection{Alterations}

The integral expressions for the unperturbed
weak rates $\lambda_{np}$ and $\lambda_{pn}$
cannot be calculated in closed form, but the standard
code \cite{kawano} allows
for either numerical integration at each temperature step, or for
the use of a series approximation (in $1/T_\gamma$).
We have opted for the numerical routine.
The perturbations to the weak rates are implemented very simply:
the numerical solutions for $\delta\lambda_{np}$ and
$\delta \lambda_{pn}$ calculated in Ref. \cite{dt} are
added to the unperturbed rates by means of a look-up table.

The effect of the back reaction of neutrino heating
on the electromagnetic plasma is more complicated.
To begin, it is useful to describe how the
evolution of the photon temperature is computed.
At each time step all the energy densities and
their derivatives are computed, and then stepped
forward in time by a Runge-Kutta integrator.
The time rate of change of the photon temperature
can be written as
\begin{equation}
{dT_\gamma \over dt} = {d\ln V\over dt}\,{d T_\gamma \over d\ln V}
=3H\,{dT_\gamma \over d\ln V} .
\end{equation}
The first law can be used to calculate $dT_\gamma /d\ln V$:
\begin{equation}
{dT_\gamma \over d\ln V} = -{\rho_{\rm EM}+p_{\rm EM} +4\delta
\rho_\nu /3 \over d\rho_{\rm EM}/dT_\gamma +d\delta\rho_\nu /dT_\gamma}.
\end{equation}

Once the evolution of the photon temperature is known,
the evolution of all other quantities (light-element
abundances and so on) follows as in
the standard case.  For example, the evolution of the
baryon-to-photon ratio $\eta$ is governed by
\begin{equation}
d\ln \eta / dt = -3d\ln (RT_\gamma) /dt .
\end{equation}
Due to $e^\pm$ annihilations $RT_\gamma$ is not
constant, and $\eta$ decreases with time.

A technical note for the experts; in the nucleosynthesis
code the first-law expression for $dT_\gamma /d\ln V$ is
actually somewhat more complicated because it also takes
into account:  the slight excess of electrons over positrons
(electron chemical potential $\mu_e$ of order $10^{-10}T$),
the tiny energy density and pressure associated with
baryons, and the bookkeeping associated with nuclear-binding energies.
Since these effects are small and unaffected by neutrino heating,
we have left them out of our discussion here.

\section{Results and Conclusions}

The ``input data'' to the nucleosynthesis code needed to
compute the effect of neutrino heating on the primordial
nucleosynthesis are:  $\delta\rho_\nu /\rho_\nu$, $\delta\lambda_{pn}$,
and $\delta\lambda_{np}$.  We consider two approaches to
computing these quantities:  (I) the detailed Boltzmann
treatment where the perturbations to the neutrino phase-space
distributions are computed \cite{dt}; and
(II) the bulk heating approach, where it is assumed
that the distortions to the neutrino distributions are thermal
and only the bulk transfer of energy from $e^\pm$ annihilations
is computed \cite{hh,rm}. In the bulk-heating approach
the effect of neutrino heating is a
slight increase in the neutrino temperature; we use the results
of Ref. \cite{rm} for $\delta T_{\nu_e}$ to compute $\delta\lambda_i$.
While we feel that the first approach is more accurate,
we have also considered the bulk-heating approach because in Ref. \cite{rm}
a very large change in the $^4$He abundance is claimed,
$\Delta Y = -0.003$. We refer the reader to Refs. \cite{rm} and
\cite{dt} for details about the two approaches.

The evolution of the energy transfer from the electromagnetic plasma
to the neutrinos is shown in Fig. 1
for the two methods of computing neutrino
heating; asymptotically $\delta\rho_\nu /\rho_{0\nu}$ approaches
$7\times 10^{-3}$.  It is heartening that these two different
treatments agree within 15\% or so
on the integrated magnitude of the distortion to
the neutrino distributions.
One consequence of the energy transfer is that there are
more electron neutrinos and they have higher energies, and
so the rates for the processes that
interconvert neutrons and protons {\it increase}.
However, there is no free lunch:  The temperature
of the electromagnetic plasma drops since it loses energy to
the neutrinos \cite{explain}.  Thus a second consequence
of the energy transfer is a {\it decrease}
in the neutron-proton interconversion rates due
to the drop in the temperature
of the electrons and positrons. This is a straightforward---but very
important---implication of energy conservation.

The third consequence of the neutrino heating
is also related to the drop
in the temperature of the electromagnetic plasma.
At a fixed time, the photon temperature
is slightly lower than in the absence of
heating; equivalently, at a fixed photon
temperature the Universe is slightly younger than
in the absence of neutrino heating.  As is well appreciated, the
$^4$He abundance is determined by the neutron fraction
at the onset of nucleosynthesis ($T_{\rm nuc} \sim 0.07\MeV$);
which, in part, is determined by the number of
neutrons that have decayed by this time.  Since the Universe is slightly
younger, fewer neutrons will have decayed. We dub this the ``clock effect.''

To summarize, there are three effects:
(i) increase in neutron-proton
interconversion rates due to neutrino heating;
(ii) decrease in neutron-proton interconversion rates due to the drop
in the temperature of the electromagnetic plasma;
and (iii) the clock effect.
The discrepancy over the sign of the change
in the $^4$He abundance traces to the fact that with the
exception of Ref.~\cite{dt}, all other authors have only
considered the first of these three effects.

First, consider the change
due to the distorted electron-neutrino distribution.
For simplicity let us begin by assuming
that the perturbation is thermal (method II),
characterized by a change in the electron-neutrino
temperature $\delta T_{\nu_e}$.
The change in the neutron fraction $X_n$ at the onset
of nucleosynthesis due to a change in either the neutrino
or the temperature of the electromagnetic plasma
is found numerically to be \cite{eureka}
\begin{equation} \label{eq:dxn}
\delta X_n = - 0.1 \delta T/ T.
\end{equation}
It is easy to understand the
sign in Eq. (\ref{eq:dxn}): when the temperature rises, the
rates for neutron-proton interconversions
increase and the neutron fraction tracks its
equilibrium abundance, $X_n/(1-X_n) = \exp (-\Delta m /T) $,
longer, which leads to a lower neutron abundance
when nucleosynthesis commences.

What is $\delta T_{\nu_e}$?  Since the electron neutrinos have both
charged- and neutral-current weak interactions, they get more than their
share of the energy transferred to the neutrinos,
about as much as mu and tau neutrinos combined. Therefore,
\begin{equation} \label{eq:delt}
        {\delta T_{\nu_e} \over T_{\nu_e}}  =  {1\over 4} {\delta
        \rho_{\nu_e}\over\rho_{\nu_e}}
        \simeq {3\over 8} ~{\delta\rho_\nu\over \rho_\nu} .
\end{equation}
Fig. 1 shows that $\delta\rho_\nu/\rho_\nu \simeq 7\times 10^{-3}$,
and thus it follows that the change in the neutron fraction
due to the fact that electron neutrinos are hotter is
$\delta X_n^{\nu} = -2.6\times 10^{-4}$.

This is not the whole story; there is a change in the neutron fraction of
opposite sign due to the slight decrease in the temperature
of electrons and positrons, which we also estimate
by Eq. (\ref{eq:dxn}).  If we assume that electrons
and positrons are relativistic (a good approximation since
the neutron fraction freezes out at a temperature of
about $0.7\MeV$) and ignore
the small differences between $e^\pm$'s and
$\gamma$'s due to statistics, then
electrons, positrons, and photons each lose the
same amount of energy due to neutrino heating.
Remembering $\delta\rho_{\rm EM} =-\delta\rho_\nu$, it follows that
\begin{equation} \label{eq:delte}
{\delta T_\gamma \over T_\gamma} \simeq - {1\over 4}
        {\delta \rho_\nu \over \rho_\nu} .
\end{equation}
Comparing Eqs. (\ref{eq:delt}) and (\ref{eq:delte})
we see that the fractional change in the
electron temperature is $-2/3$ that of the electron-neutrino
temperature, leading to
an {\it increase} in the neutron fraction that is only
2/3 as large, $\delta X_n^\gamma \simeq 1.7\times 10^{-4}$.
The predicted net change in the neutron fraction is thus
\begin{equation} \label{eq:bulk}
\delta X_n \equiv \delta X_n^\nu + \delta X_n^\gamma \simeq - 0.1
        \left( {3\over 8} - {1\over 4} \right)
        \,{\delta\rho_\nu \over \rho_\nu } \simeq -9\times 10^{-5}.
\end{equation}
Figure 2 shows $\delta X_n$ as a function of temperature;
the numerical results agree well with this simple analytical
prediction.

Figure 2 also shows the result of incorporating neutrino heating
into the code via method I, where the
distortion is not assumed to be thermal.  In fact,
as discussed in Ref. \cite{dt}, the perturbation
to the neutrino spectra is highly nonthermal
due to the fact that more high-energy neutrinos are
produced in the process of neutrino heating
since neutrino cross sections rise with energy.
This excess of high-energy neutrinos further
enhances the neutron-production rate, which as the temperature
drops is becoming more suppressed
by the neutron-proton mass difference, and
therefore we expect $\delta X_n^\nu$
to be larger---which is precisely what is seen in
Fig.~2.  Since $\delta X_n^\nu$ is larger, the near
cancellation between $\delta X_n^\nu$ and $\delta X_n^\gamma$
is even more precise:  For method I, $\delta X_n
\simeq -2\times 10^{-5}$.

For reference, the the mass fraction of $^4$He synthesized is related
to the neutron fraction at freeze out by:  $Y\simeq
1.33X_n$ (see e.g., Ref \cite{dt}).  Thus,
the predicted change in the $^4$He mass
fraction due to the first two effects is:
$\Delta Y_{1+2} \simeq -3\times 10^{-5}$ (method I) and
$-1.1\times 10^{-4}$ (method II).

The clock effect involves the age of the Universe at the epoch at
which nucleosynthesis commences,
$T=T_{\rm nuc} \simeq0.07\MeV$.   Since the Universe is slightly younger
when nucleosynthesis commences when neutrino heating
is taken into account, fewer neutrons decay from the time
that the neutron fraction freezes out, leading
to a larger $^4$He abundance.  In Ref. \cite{dt} the
change in the $^4$He abundance due to the clock effect
was estimated to be $\Delta Y_{\rm clock} \simeq
+1.5\times 10^{-4}$.  Figure 3 shows the total change
in $^4$He abundance as computed by our modified version
of the standard code.  For method I, $\Delta Y$ is
about $+1.5\times 10^{-4}$, while for method II it
is about $+1.1\times 10^{-4}$, which indicates
that the $\Delta Y_{\rm clock} \sim 2\times 10^{-4}$,
in reasonable accord with the previous estimate.

There are a couple of fine points to be made about the
baryon-to-photon ratio.   In the standard scenario
the baryon-to-photon ratio decreases by a factor
of 4/11 from its pre-nucleosynthesis value
to its post-nucleosynthesis value,
due to the entropy transfer from $e^\pm$ pairs
to the photons (see Fig.~4).  When neutrino heating
is taken into account the decrease is less,
by about 0.5\% (see Fig.~4), which means that
for a {\it fixed value of $\eta$ today, the value
of $\eta$ before $e^\pm$ annihilations was smaller.}
We remind the reader that one always specifies
the yields of primordial nucleosynthesis
in terms of the present value
of the baryon-to-photon ratio.   This suggests
a fourth effect of neutrino heating on nucleosynthesis,
involving the fact that the value of $\eta$ at early times is always
smaller when neutrino heating is taken into
account; we dub this the $\eta$-effect \cite{dt}.
Thankfully, this effect for most values of $\eta$
is small because somewhat
before the onset of nucleosynthesis $\eta$ has
reached its asymptotic (present) value; see Fig.~4.
For the most interesting values,
$10^{-10}\la \eta \la 10^{-9}$, $\Delta Y$ is
insensitive to the value of $\eta$.  For extreme
values of $\eta$ it becomes $\eta$ dependent.

For large values of $\eta$, $\Delta Y$ depends
upon $\eta$ because of the $\eta$-effect:  As one increases
$\eta$ the onset of nucleosynthesis occurs earlier;
for large enough $\eta$ it occurs before
$\eta$ has reached its asymptotic value; thus,
the value of $\eta$ during nucleosynthesis is slightly
{\it smaller} when neutrino heating is taken into account;
since $Y$ increases monotonically with $\eta$,
the amount of $^4$He synthesized decreases due to
this effect.  This is precisely the behaviour seen
in Fig.~3:  For $\eta \gg 10^{-9}$, $\Delta Y$ decreases.

To understand why $\Delta Y$ also decreases
for very small values of $\eta$, we must first recall why
the primordial helium abundance drops so
precipitously for small values of $\eta$ (for $\eta\la
10^{-11}$ the mass fraction of D synthesized is
actually greater than that of $^4$He).  Small
$\eta$ means that number densities of all nuclear species are
small, so that nuclear-reaction rates are correspondingly
lower:  $\Gamma_{\rm nuclear} \propto \eta$.
For extremely low values of $\eta$, by the time
nucleosynthesis commences nuclear reaction rates
have been come ineffective ($\Gamma_{\rm nuclear}
\la H$), and the amount of $^4$He produced depends
upon the relative effectiveness of the nuclear
reactions:  $Y\propto \Gamma_{\rm nuclear}/H$.
Neutrino heating increases the expansion rate
(at fixed photon temperature), therefore
the $\Gamma_{\rm nuclear}/H$ is {\it smaller} and
less $^4$He is synthesized.  This is precisely
what is seen in Fig.~3:  For $\eta \ll 10^{-10}$,
$\Delta Y$ decreases with decreasing $\eta$.

To conclude, neutrino
heating affects the synthesis of $^4$He in
four distinct ways; by incorporating the effect of the
slight heating of neutrinos by $e^\pm$ annihilations
into the standard nucleosynthesis code we have
quantified its effect on nucleosynthesis and
clarified previous conflicting estimates.
The net result of the four effects is a slight
increase in the mass fraction of $^4$He synthesized,
$\Delta Y \simeq +1.5\times 10^{-4}$, for
the interesting range of $\eta$.

\vskip 1.5cm
\noindent  We thank David Thomas and David Schramm
for useful discussions.  This work was supported in part by the
DOE (at Chicago and Fermilab) and by the NASA through
NAGW-2381 (at Fermilab).

\newpage

\begin{center}
{\bf FIGURE CAPTIONS}
\end{center}
\medskip

\noindent {\bf Figure 1:}  The evolution of the total perturbation to the
neutrino density due to heating by $e^\pm$ annihilations as
a function of the photon temperature.  The solid curves are the results
from Ref. \cite{dt}; the broken curves are those from Ref. \cite{rm}.

\medskip
\noindent {\bf Figure 2:}  (a)  The evolution of the
neutron fraction $X_n$ as a function of
the photon temperature:  $X_n$ tracks its equilibrium
value until $T_\gamma \sim 0.3\MeV$, when it levels off because
of the freeze out of the weak interactions; it then
slowly decreases due to neutron decays; the precipitous drop occurs
because of the onset of nucleosynthesis ($T_\gamma\sim 0.07\MeV$).
(b) The change in the neutron fraction $\delta X_n$ due to the
effects of neutrino heating as a function
of the photon temperature:  $\delta X_n$ begins to level
off at $T_\gamma\sim 0.2\MeV$ due to the freeze out of the weak interactions;
it then rises because at a given value of $T_\gamma$ the
Universe is younger and fewer neutrons have decayed (``clock effect'');
it drops to zero when nucleosynthesis commences.
The solid curves are based upon the results of Ref. \cite{dt};
the broken curves upon those of Ref. \cite{rm}; these results
are for $\eta = 3\times 10^{-10}$.

\medskip
\noindent {\bf Figure 3:}  The change in the predicted
$^4$He abundance due to neutrino heating as a function
of the present baryon-to-photon ratio.
The solid curves are based upon the results of Ref. \cite{dt};
the broken curves upon those of Ref. \cite{rm}.

\medskip
\noindent {\bf Figure 4:}  (a)  The evolution of the baryon-to-photon
ratio as a function of the photon temperature; (b)  The
in change baryon-to-photon ratio, $\Delta\eta /\eta$,
due to neutrino heating as a function of photon temperature.
Note, we have chosen the initial value of $\eta$ with and
without neutrino heating so that the final value is identical.
The solid curves are based upon the results of Ref. \cite{dt};
the broken curves upon those of Ref. \cite{rm}.


\vskip 2 cm
\begin{thebibliography}  {rates}

\bibitem{ytsso} J.~Yang, M.S.~Turner, G.~Steigman, D.N.~Schramm,
and K.A.~Olive, {\it Astrophys. J.} {\bf 281}, 493 (1984).

\bibitem{nulimit} V.F.~Shvartsman, {\it JETP Lett.} {\bf 9},
184 (1969); G.~Steigman, D.N.~Schramm, and J.~Gunn, {\it Phys.
Lett. B} {\bf 66}, 202 (1977).

\bibitem{rates} M.S.~Smith, L.H.~Kawano, and R.A.~Malaney,
CITA preprint OAP-716 (1992).

\bibitem{dicusetal} D.~Dicus et al., {\it Phys. Rev. D}
{\bf 26}, 2694 (1982).

\bibitem{matthews} G.~Mathews and R.A.~Malaney, {\it Phys.
Repts.}, in press (1992).

\bibitem{lithium} M.~Spite and F.~Spite, {\it Nature}
{\bf 297}, 483 (1982); M.~Spite, J.P.~Maillard, and F.
Spite, {\it Astron. Astrophys.} {\bf 141}, 56 (1984);
R.~Rebolo, P.~Molaro, and J.~Beckman, {\it ibid}
{\bf 192}, 192 (1988); L.~Hobbs and J.~Thorburn,
{\it Astrophys. J.} {\bf 375}, 116 (1991); K.~Olive and D.N.~Schramm,
{\it Nature}, in press (1992); J.~Thorburn, in preparation (1992).

\bibitem{dhe3} J.~Yang et al., {\it Astrophys. J.} {\bf 281}, 493 (1984);
D.S.P.~Dearborn, D.N.~Schramm, and G.~Steigman, {\it Astrophys. J.}
{\bf 302}, 35 (1986); T.M.~Bania, R.T.~Rood, and T.L.~Wilson,
{\it ibid} {\bf 323}, 30 (1987); J.~Linsky, {\it Astrophys. J.},
in press (1992).

\bibitem{he4} B.E.J.~Pagel, {\it Physica Scripta} {\bf T36},
7 (1991); B.E.J.~Pagel and A.~Kazlauskas, {\it Mon. Not. R. astron. Soc.},
in press (1992).

\bibitem{walker} T.P.~Walker et al., {\it Astrophys. J.} {\bf 376},
51 (1991).

\bibitem{Weinberg} S.~Weinberg, {\it Gravitation and Cosmology}
(J.~Wiley, NY, 1972), Ch.~15.

\bibitem{hh} M.A.~Herrera and S.~Hacyan, {\it Astrophys. J.}
{\bf 336}, 539 (1989).

\bibitem{rm} N.C.~Rana and B.~Mitra, {\it Phys. Rev. D}
{\bf 44}, 393 (1991).

\bibitem{df} A.D.~Dolgov and M.~Fukugita, {\it Phys. Rev. D},
in press (1993).

\bibitem{dt} S.~Dodelson and M.S.~Turner, {\it Phys. Rev. D},
in press (1992).

\bibitem{kawano} L.~Kawano, Fermilab Pub 92/04-A, (1992).

\bibitem{explain} In Ref. \cite{dt} the back reaction
of neutrino heating on the electromagnetic plasma
is carefully discussed; at fixed value of the cosmic
scale factor, $T_\gamma$ is lower when neutrino heating
is taken into account by $\delta T_\gamma
= -\delta\rho_\nu /(d\rho_{\rm EM}/dT_\gamma )\approx
-{1\over 4}(\delta\rho_\nu /\rho_{0\nu})T_\gamma$, which
follows by energy conservation.

\bibitem{eureka}  The empirical relation $\delta X_n
\simeq -0.1 \delta T_i /T_i$ can be derived.  To do
so, define all perturbations with respect to fixed
value of the scale factor, cf. Ref. \cite{dt}.  The
freeze-out value of $X_n$ is set by its equilibrium
value when $\Gamma /H \sim 1$:  $X_n /(1-X_n)=
\exp (-\Delta m/T_f)$, where $(\Gamma /H)|_{T_f} =1$.
Write $\Gamma =aR^{-5}(r_\nu^5+r_e^5)$ and $H=bR^{-2}$,
where $R^{-1}=T_{0\nu}$ (neutrino temperature
in absence of $e^\pm$ heating), $r_\nu =T_\nu/T_{0\nu}\sim 1$,
$r_e=T_\gamma /T_{0\nu}\sim 1$.  The effective
temperature ``felt'' by nucleons, which determines
the equilibrium neutron-to-proton ratio,
$T_{\rm eff} \sim (T_\nu +T_\gamma )/2 = R^{-1}(x_\nu +x_e)$.
Since $\delta\rho_{\rm EM} = -\delta\rho_\nu$ neutrino
heating does not affect the expansion rate.  After some
algebra it follows that $\delta X_n = -{1\over 3}
X_n(1-X_n)(\Delta m/T_f)\delta T_i/T_i \approx
-0.1\delta T_i/T_i$.

\end{thebibliography}
\end{document}